\documentclass[twocolumn,showpacs,preprintnumbers,amsmath,amssymb]{revtex4}

\usepackage{graphicx}

\usepackage{bm}

\begin{document}

\newcommand{\non}{\nonumber}
\newcommand{\be}{\begin{equation}}
\newcommand{\ee}{\end{equation}}
\newcommand{\bq}{\begin{eqnarray}}
\newcommand{\eq}{\end{eqnarray}}
\newcommand{\bsp}{\begin{split}}
\newcommand{\esp}{\end{split}}
\newcommand{\lps}{\langle}
\newcommand{\rps}{\rangle}

\title{Nonadiabatic effects in the dynamics of atoms confined in a cylindric 
time-orbiting-potential magnetic trap}

\date{\today}

\author{Roberto Franzosi} 
\email{Roberto.Franzosi@df.unipi.it}
\affiliation{INFN Sez. di Pisa and INFM UdR di Pisa, Dipartimento di 
Fisica E. Fermi Universit\`a di Pisa,
Via Buonarroti 2, I-56127 Pisa, Italy.}
\author{Andrea Spinelli} 
\author{Bruno Zambon} 
\email{Bruno.Zambon@df.unipi.it}
\author{Ennio Arimondo} 
\affiliation{INFM UdR di Pisa, Dipartimento di Fisica E. Fermi 
Universit\`a di Pisa,
Via Buonarroti 2, I-56127 Pisa, Italy.}

\begin{abstract}

In a time-orbiting-potential magnetic trap the neutral atoms are 
confined by means of an inhomogeneous magnetic
field superimposed to an uniform rotating one.  We perform an analytic
study of the atomic motion  by taking  into account  the nonadiabatic
effects arising from the spin dynamics about the local magnetic field. 
Geometric-like magnetic-fields determined by the Berry's
phase appear within the quantum description. The application of a 
variational procedure  on the original quantum equation leads to a set of 
dynamical evolution equations  for the quantum average value of 
the position  operator and of the spin variables. 
Within this approximation we derive the quantum-mechanical ground state 
configuration matching the classical adiabatic solution and perform some 
numerical simulations.

\end{abstract}
\maketitle

\section{Introduction}
The difficulty of analysing a complex physical system is  greatly reduced 
when one is able to identify a certain number of different time scales present
in the system  dynamical evolution. Thus,   a series of approximations, 
generically termed adiabatic approximations, can successfully  be carried out.  
The very simple basic idea is that of  dealing first  with the motion 
of the fast variables, keeping the slow ones fixed but arbitrary, and then to
complete the analysis of the entire  system by allowing a
variation of the previously fixed coordinates. The quantum
adiabatic theorem and the molecular Born-Oppenheimer approximation
are well-known examples of this approach, with its origins in the
early days of quantum mechanics. The quantum-adiabatic theorem
dictates that a system prepared in an eigenstate of its
Hamiltonian will remain in the corresponding eigenstate as the
Hamiltonian is varied slowly enough. If the Hamiltonian returns 
to its original form, the
system assumes the original eigenstate multiplied by an
appropriate dynamical phase factor related to the instantaneous 
eigenvalue of the Hamiltonian. Berry
made the interesting observation that in addition to the dynamical
phase factor produced by the eigenvalue time  evolution, the
wavefunction acquires an additional phase
contribution~\cite{wilczek}. This additional contribution, the
geometric phase, depends only on the path travelled by the system in
the space of external parameters.\\
\indent A canonical example for a  system where this behavior occurs is that
of a neutral particle carrying a magnetic moment and moving in an
inhomogeneous magnetic field. Here, the fast variable is the
transverse magnetic moment, and the slow variable is the atomic
position and momentum \cite{majorana}. If the magnetic field
varies slowly enough in space,  the effective
Hamiltonian governing the dynamics of the slow external variables
contains an induced gauge potential, the so-called geometric
potential. In the classical limit the  gauge geometric fields
acting on the neutral particle with a magnetic moment have been
studied by Aharonov and Stern~\cite{aharonov}; they found that the atom 
experiences geometric  Lorentz-type
and electric-type forces \cite{littlejohn}. The
magnitudes of these forces do not depend on the amplitude of the 
magnetic field, but only 
on its local orientation.\\ 
\indent In order to treat the non-adiabatic corrections improved 
Born-Oppenheimer 
methods were introduced for the case of arbitrary spin values \cite{sun}. 
Later the non-adiabatic terms modifying the atomic motion  have been 
studied by several authors in the context of magnetic structures guiding 
or confining very cold atoms \cite{ho,sukumar,hinds00,potvliege}. 
The non-adiabatic corrections produce  
spin-flip transitions leading to atomic loss from the magnetic 
configurations, and also they modify the atomic motion.  High order
post-adiabatic corrections, leading to geometric
electromagnetism potentials, have been investigated
for the elegant configuration of an atom orbiting around a
straight current-carrying wire \cite{schmiedmayer}. For the verification
of Berry's phase and its consequences a natural question is whether
one can observe the direct modification of the atomic motion in the
classical limit for the induced gauge potentials.  Measurements on the
motion of a rubidium Bose-Einstein condensate in a
time-orbiting-potential (TOP) magnetic trap represent a quite strong 
indication for the existence of these  geometric forces~\cite{muller00,cr00}.
  Those
observations were analyzed through a classical description for the
condensate center-of-mass motion and for the atomic magnetic moment.\\ 
\indent In the present paper we perform an analytic study of the atomic
quantum dynamics within TOP magnetic traps.  We take into account 
the non-adiabatic effects  arising from the dynamics of the spin orientation 
around the local magnetic field. Within a pure quantum description, the 
geometric magnetic fields appear as a consequence of the presence
of inhomogeneous magnetic fields. In  this context spinors quantities 
can be  introduced to
describe the atomic spin states, as done by Ho and
Shenoy~\cite{ho} for the Berry phase in atomic 
condensates in magnetic traps. We derive an effective atomic dynamics, by
means of a time-dependent variation principle making the
quantum description analytically treatable.  Thus, the atomic
motion results by the coupling of a quantum harmonic motion, governing the 
atomic
scalar wave function, and an effective nonlinear spin dynamics. 
The harmonic Hamiltonian depends on time-varying parameters that, in
turns, are linked to the spin state.  Also the spin Hamiltonian
parameters are time-dependent, as they result from  the atomic 
wave-functions 
expectation values of the above geometric operators.  Within this
non-adiabatic approximation the ground state configuration matches
well the adiabatic solution. We also performed numerical simulations
with these new equations. For the parameters suggested by the standard 
experimental set-up we found that the adiabatic approximation is well
suited. Nevertheless, by reducing the intensity of the bias field, 
non-adiabatic effects show up, because under these conditions the
influence of the geometric fields is more relevant.\\
\indent Section \ref{toptrap} summarizes the classical analysis based 
on the adiabatic approximation and leading to 
the atomic micromotion.
Section \ref{adiabatic} reports a quantum analysis of the atomic motion
within the adiabatic approximation by taking into account the lowest 
frequency 
terms of the  time-dependent potential. 
Within this approximation we recover the quantum counterpart of the 
classical micromotion.
Section \ref{Quantum} studies the quantum dynamics
of atoms into TOP traps.
In Section~\ref{Beyond},  by means of a time-dependent-variation-principle, 
we derive an 
effective dynamics for the atomic motion.
Section~\ref{Numerical} reports numerical simulations for the dynamical regime
of the atomic motion.


\section{TOP trap}
\label{toptrap}
\begin{figure}[ht]
\includegraphics[scale=0.5]{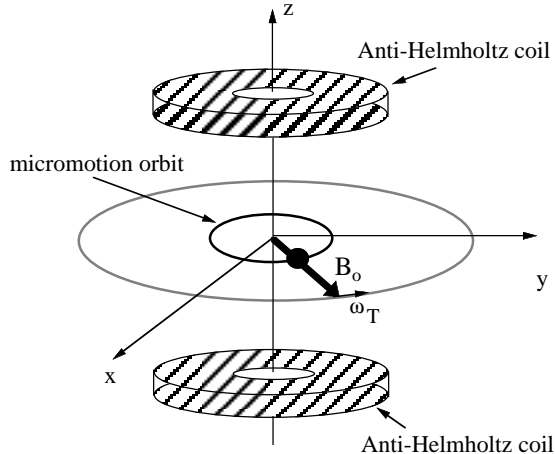}
\caption{Schematic representation of a cylindric TOP trap with the 
anti-Helmholtz coils producing a quadrupole field with vertical 
symmetry, the $B_{0}$ bias field rotating at angular frequency 
$\omega_{\rm T}$ in the horizontal plane, and the atomic cloud (black 
sphere) following the micromotion orbit.}
\label{setup}
\end{figure}

Bose-Einstein condensation in dilute atomic gas is created  by trapping
cold atoms in a magnetic trap of which the Ioffe-Pritchard (IP) and the 
time-orbiting-potential  are the most common  ones.  In a TOP trap 
the magnetic field, schematically represented in the inset of 
Fig.~\ref{setup}, is
composed of a quadrupole (inhomogeneous) field  and a rotating
(time-dependent) bias field, $B_0$. The TOP trap, introduced in~\cite{petrich} 
for the very first experiments on Bose-Einstein
condensation~\cite{cornell95}, is  employed by a
number of research groups producing Bose-Einstein 
condensates~\cite{hagley,anderson,han,martin,arlt}. \\
\indent The single particle   Hamiltonian for the  atoms inside 
a magnetic field configuration which characterizes   the trap geometry
is given by 
\be
H(t) = \frac{{\bf p}^2}{2m} + m g z - \frac{\mu}{s} {\bf s}\cdot 
{\bf B}({\bf x},t) 
\label{genHam}
\ee
where ${\bf s}$ are the spin operators of the $s =\hbar j $ representation, the
last term takes into account the magnetic interaction energy of an atom 
with magnetic moment $\mu {\bf s}/s$ and projection $\mu=-|\mu|$ along 
the magnetic field ${\bf B}$.  We also adopt the representation
${\bf x}= x \hat{x} +  y \hat{y} +  z \hat{z}$ for the position vector. 
For a TOP trap the magnetic field is the superposition of a static
quadrupole field and one rotating at the radiofrequency (RF) $\omega_T$ 
\be
{\bf B}({\bf x},t) = {\bf b}({\bf x}) + {\bf B}_t(t) \, .
\label{totalB}
\ee
Its components are 
\be {\bf b}({\bf x})= b_x x \hat{x} + b_y y \hat{y} 
+ b_z z \hat{z} \, .
\label{gradient}
\ee
and
\be {\bf B}_t(t) = B_{0} \cos (\omega_T t) \hat{x} + B_{0} \sin 
(\omega_T t) \hat{y} \, .
\label{rotating}
\ee 
The magnetic field parameters define the specific type of TOP 
trap we are analyzing \cite{note}. By supposing the RF field rotating in the 
horizontal ${x,y}$ plane we define the TOP geometry of the 
traps operating at Boulder~\cite{petrich} and at Pisa~\cite{jphysbpaper}. 
In this work
we analyze the dynamics of a cylindric TOP trap with 
\be
b_x=b_y=-b_z/2=b \, .
\label{ciltop}
\ee
\\
For the usual TOP trap three different time scales exists:
the  fastest motion, given by the Larmor frequency $\omega_L$ and   related 
to the spin precession around the local magnetic field;  
the magnetic bias field rotating  frequency $\omega_T$  giving
rise to time-dependent forces at the frequences $\omega_T, 2 \omega_T,\ldots$
and a slower motion  associated to the atom spatial motion given
by the trap harmonic frequency $\omega_{h}$. These time scales 
are in order of magnitude, i.e.  $\omega_h << \omega_T << \omega_L$ thus
making possible the adiabatic approximation.  In the first place the
fast spin precession around the local magnetic field allows to
consider the atoms spin locked to the local magnetic field throughout
the whole spatial motion.  This leads to an adiabatic time dependent
potential $U=-{\mu \over s} {\bf s}\cdot {\bf B}({\bf x},t) = |\mu|
\cdot |{\bf B}({\bf x},t)|$.  Secondly by averaging in time (over a
period $2 \pi/\omega_T$) this potential and by keeping only the
slowest component gives rise to a harmonic potential spatially
confining the condensate.  The next order of approximation examines
the fast variables (fast with respect to the harmonic dynamics at
frequency $\omega_h$) related to the time-dependent potential (at
frequency $\omega_T$).  An exhaustive computation of this
approximation is found in~\cite{Minogin}.  \\
\indent Gov {\it et al}~\cite{GovStr} have used the standard   classical equations 
of the atomic motion of a  magnetic moment within a TOP  in which all the 
different time scales are present. In this context a steady periodic 
orbit can be found exactly without resorting to any approximation. This 
periodic solution corresponds to  what it is known as the {\it atomic 
micromotion} \cite{SPIE} (see Fig.~\ref{setup}). As the atoms trace out the 
atomic micromotion orbit, the magnitude and 
direction of the local magnetic field change in space and time, with the 
magnetic moment of the atom precessing around the direction of the field.
In fact, given the Hamiltonian of Eq. (\ref{genHam}) 
a dynamical state
$|\Psi \rps$ satisfies the following center-\-of-\-mass 
equations of motion:
\be
\begin{split}
m \frac{d^2{\bf R}}{dt^2} &= \frac{\mu}{s} \nabla [{\bf S} 
\cdot {\bf b}({\bf R}) ]
- m g \hat{z} \\
\frac{d{\bf S}}{dt} &= \frac{\mu}{s} \lps \Psi |{\bf s} \wedge 
{\bf B}({\bf x},t) | \Psi\rps \, ,
\end{split}
\ee
where ${\bf R} = X\hat{x}+ Y\hat{y}+ Z\hat{z} = \lps 
\Psi |{\bf x} | \Psi\rps$ is the expectation value of the center 
of mass position, and 
${\bf S} = S_x\hat{x}+ S_y\hat{y}+ S_z\hat{z} =
\lps \Psi |{\bf s} | \Psi\rps$ is the expectation value of the atomic spin. 

Thus we are lead to a set of equations which are   not closed. However  
 if the quantum mechanical wave
function can be factorized, as to the lowest order approximation,
we may write $\lps {\bf s} \wedge {\bf B}({\bf x},t) \rps = \lps {\bf s} \rps
\wedge \lps{\bf B}({\bf x},t) \rps$ .  Within this approximation the
above system of equations becomes closed and assumes the form \be
\begin{split}
m \frac{d^2{\bf R}}{dt^2} &= \frac{\mu}{s} \nabla [{\bf S} 
\cdot {\bf b}({\bf R}) ]
- m g \hat{z} \\
\frac{d{\bf S}}{dt} &= \frac{\mu}{s} {\bf S} \wedge 
{\bf B}({\bf R},t)  \, .
\end{split}
\label{comeom}
\ee
The simplest periodic solutions generated by the Eqs. (\ref{comeom})
give a good estimate of the fast center-\-of-\-mass condensate  
dynamics, i.e. the atomic micromotion. This motion  was experimentally observed
in a triaxial TOP trap in Ref.~\cite{muller00}. This periodic solution 
is best viewed in a frame rotating with the bias magnetic field. In this 
frame the magnetic
moment results to be aligned to the effective magnetic field   and the 
atomic center of mass is at rest. Also, in order to have a stable 
motion, the spin must be tuned to the effective magnetic field,
producing a confining potential energy.  The stability along the $z$ axis
requires a zero force along this direction: this leads to \be {S_z
\over s}={{ m \, g} \over {2 b |\mu|}}=\xi \ee which accounts for the
fact that the component of the spin along $z$ must be positive.  The
centrifugal force balancing the gradient force yields the radius of
the micromotion \be r = \frac{|\mu| b }{m \omega^2_{T}} \alpha \ \ \ \
\ \ \ \ \alpha = \sqrt{1-\xi^{2}}
\label{clr}
\ee
where $\xi$ and $\alpha$ are respectively the cosinus and sinus that 
the effective field ${\bf B}^{inst}$ forms with the $z$ axis.
The components of the effective field in the rotating frame, with 
horizontal component along the $x$ axis, are 
\be 
B_x^{inst}= b \, r  \ \ \ \ \ \ {\rm and } \ \ \ \ \ B_z^{inst}=-2\, b\, z 
+ \omega_T s/ \mu.
\label{Binst}
\ee
In order to determine the  $z$ height of the periodic orbit 
under examination, the following parameters  also useful
in the subsequent analysis are required: 
\be \varsigma = \frac{\omega_T s}{\mu b},
\quad \rho = \frac{B_{0}}{b}\, .  
\label{par1}
\ee
$\rho$ being the radius of the circle of death~\cite{petrich} and 
$\varsigma$ is twice the amount the zero point of the quadrupole field 
shifts downwards for effect of  the uniform fictitious magnetic  field 
which appears in the rotating frame.  In terms of these parameters 
the equilibrium height can be expressed by means of the following 
alignment relation 
\be 
{{-2\, b \, (z - {\varsigma \over 2}) }\over { B_0 + b \, r}}= 
{ \xi \over \sqrt{1 -\xi^2}}
\ee
which leads to 
\be 
z={\varsigma \over 2} - {\xi \over \sqrt{1 -\xi^2}}
({ {B_0 + b\, r} \over {2 b} })= {\varsigma \over 2}- {\xi \over 2} 
( {\rho \over \alpha} + {{ |\mu| b}\over { m \, \omega_T^2}} ) 
\label{zeq}
\ee

The above analysis  suggest that the adiabatic approximation is
more rigorous if we refer to ${\bf B}^{inst}$ instead of the
real magnetic field. Indeed, in  the  motion described above the adiabatic 
approximation is completely fulfilled with respect to this field. However, 
for more general solutions of the Eqs. (\ref{comeom}) the behavior 
is different   and we must also account 
for  a spin component orthogonal to the local magnetic field. In any case, 
as it will be shown in the numerical simulations,  
the projection of the magnetic moment along this field is a  much better  
conserved quantity with respect to  that  along the real field. 
In Refs.~\cite{aharonov,levitron} the effect of the  components 
perpendicular to the real magnetic field  have  been 
examined and it was found that 
a small misalignment with respect to this  
field gives rise to a Lorentz-type force. An additional 
electric-type force is originated  
by the time average of the fast oscillatory force induced by
the spin precession. Both kind of forces, affecting 
the center-of mass dynamics, are geometric forces because they do not 
depend on the magnitude of the magnetic field but only on its orientation.

\section{From 
classical to quantum adiabatic approximation}
\label{adiabatic}
The gross features of the atomic confinement in a magnetic trap are
explained in terms of the adiabatic approximation. If this is 
fulfilled, 
an atom with electron magnetic moment parallel to the local magnetic
field experiences a confining potential given by
$U= |\mu {\bf B}|$. Working in the rotating frame 
 one replaces ${\bf B}$ with ${\bf B}^{inst}$.
Since in the following we will make use of an adiabatic-like solution
in order to find non-adiabatic corrections, we will assume the spin to form 
an angle with ${\bf B}^{inst}$, whence we will take the spin component
in the direction of the instantaneous magnetic field to be $s \sigma = {\bf s}
\cdot {\bf B}^{inst}/|{\bf B}^{inst}|$.
Thus, from Eq. (\ref{genHam}) we obtain the adiabatic Hamiltonian to be
\be
H^\sigma_{ad} = \frac{{\bf p}^2}{2m} + U_{ad} \, ,
\label{Hadiab}
\ee
where
\be
\begin{split} 
    U_{ad} &= m g z 
	+ \sigma|\mu| \Big[ (b x + B_0 \cos\omega_T t)^2 + \Big. \, \\ 
	&\Big. (b y + B_0 \sin\omega_T t)^2 + (-2b z + \frac{\omega_T s}{\mu})^2 
	\Big]^{1/2}
	\, .
\label{Udiab}
\end{split}
\ee 
 For small displacements of the atoms from the
equilibrium position $(0,0,h)$, the adiabatic potential $U_{ad}$ can
be expanded in a power series of the displacement coordinates
$(x,y,\zeta = z-h)$, and up to the second order we have 
\be
\begin{split}
U^{(2)}_{ad} =& 
 + m g \left( 1 + \frac{\sigma \eta}{\beta \xi} \right) \zeta + \, \\
&\frac{1}{2}m
\left[ \omega^{2}_{0,r}  ( x^2 + y^2) + 
\omega^{2}_{0,z}\zeta^{2} \right]
  + U_0 + U(t) \, .
\label{u2t}
\end{split}
\ee
Here the time-independent component is 
\be
U_{0} = m g h + \sigma \beta |\mu| B_0  \, ,
\ee
and the time-dependent component is 
\be
\begin{split}
U(t,\sigma) &= 
\frac{\sigma |\mu| b}{\beta} \left( 1 - 
\frac{2 \eta}{\beta^2} \frac{\zeta}{\rho} \right) \left( x \cos\omega_T t +
y \sin\omega_T t \right) -\\
& \frac{\sigma |\mu| b}{4 \rho \beta^3}
\left[ (x^2-y^2) \cos(2 \omega_T t) 
+ 2 x y \sin(2 \omega_T t) \right]
\, .
\end{split}
\ee
The new adimensional constants here introduced are 
\[
\eta = \frac{2h - \varsigma}{\rho}\, , \quad  \beta = \sqrt{1+\eta^2} \, ,
\]
while the oscillation frequencies are
\be
\omega_{0,r} = \sqrt{\frac{\sigma |\mu| b(2 \eta^2 + 1)}
{2 m \rho \beta^{3}}} \, , \quad 
\omega_{0,z} = \sqrt{\frac{4\sigma |\mu| b}
{ m \rho \beta^{3}}} \, .
\label{frequencies}
\ee
Two time scales are involved in $U^{(2)}_{ad}$, the slower one being
associated to the harmonic motion at the 
oscillation frequencies $\omega_{0,r}$ and $\omega_{0,z}$, whereas the
faster one is associated with the bias frequency $\omega_T$.
The oscillating forces have vanishing time average over a period
$2\pi/\omega_T$.
This can be substantiated by the following wave function factorization
\be
\Psi ({\bf x},t,\sigma) = \Phi ({\bf x},t,\sigma) {\cal E}({\bf 
x},t,\sigma) 
\label{wf}
\ee
where 
\be {\cal E} = \exp [ -i w /\hbar ],
\label{expfact}
\ee
and $ w= 
\int^t_0 dt U(t,\sigma) $ describes the dominant 
effects of the oscillating potential. 
The time-scale separation allows us to consider  $\Phi ({\bf x},t)$ as
 a slowly varying function of time~\cite{cook}.  Notice the
explicit dependence on the parameter $\sigma$.  Substituting Eq. 
(\ref{wf}) into the Schr\"odinger equation with the
potential $U^{(2)}_{ad}$, we get 
\be
\begin{split}
i \hbar \partial_t \Phi ({\bf x},t,\sigma) = & 
\Bigg[ 
\frac{{\bf p}^2}{2m} + U_0 + (m g  + \frac{\sigma \eta}{\beta \xi}) 
\zeta  +  \Bigg. \\ \Bigg.
&\frac{1}{2}m\left[ \omega^{2}_{0,r}  ( x^2 + y^2) + 
\omega^{2}_{0,z} \zeta^2 
\right]\Bigg] \Phi ({\bf x},t,\sigma) + \\
& \left[
\frac{i \hbar}{m} \nabla w \cdot \nabla + \frac{1}{2m }
\vert \nabla w \vert^2 + \frac{i \hbar}{2 m} \nabla^2 w
\right] \Phi ({\bf x},t,\sigma) \, ,
\end{split}
\label{Sered}
\ee
The above assumptions on the different timescales allow us to consider the coefficients of the
oscillating terms at frequencies $\omega_T$ and $2\omega_T$  as slowly varying ones. 
Indeed a time average 
over the short time $2\pi/\omega_T$ leads to
\be
\begin{split}
i \hbar \partial_t \Phi ({\bf x},t,\sigma) &=  
\Bigg\{ 
\frac{{\bf p}^2}{2m}  
+ m g\left (1  + \frac{\sigma \eta}{\beta \xi} 
- \frac{\sigma^2 \eta }{2 \rho  \beta^4 \xi^2} 
\frac{g}{\omega^2_T}  \right) \zeta + \\ 
& \frac{1}{2} m 
\left[
\omega^2_r (x^2+y^2) + \omega^2_z \zeta^2
\right] 
\Bigg\} \Phi ({\bf x},t) \, ,
\end{split}
\label{Sered2}
\ee
where the irrelevant constant terms have been dropped. 
This equation displays a three-dimensional harmonic 
oscillator structure whose frequences are  
\be
\begin{split}
\omega_r &= \omega_{0,r} \left[ 
1  +  \frac{\sigma}{16 \xi \beta^3}\frac{g}{\rho\omega^2_T} 
 \frac{(32 \eta^2 + 1)}{(2 \eta^2 + 1 )} \right]^{1/2} \, , \\ 
\omega_z &= \omega_{0,z}
\left[ 
1 + 
\frac{\sigma}{2 \beta^3} \frac{g}{\rho\omega^2_T} \eta^{2}
\right]^{1/2} \,
\end{split}
\label{corrfreqs} 
\ee
in the $xy$-plane and along the $z$-direction, respectively.
The equilibrium position along the $z$-axis is obtained by setting to zero the 
term multiplying $\zeta$ in Eq. (\ref{Sered2}).
Neglecting smaller  contributions, the equilibrium atomic position is 
given by
\be 
h(\sigma) = \varsigma/2 - \rho \xi / (2 \sqrt{\sigma^2 -
\xi^2}) \, .  
\label{height}
\ee 
In the limit $\sigma =1$ the 
classical solution of Eq. (\ref{zeq}) up to a small term containing
$\omega^{-2}_T$ is matched.  The expectation value of the
particle momentum ${\bf p}$ on the state (\ref{wf}) is
\be 
\lps {\bf p} \rps = \frac{|\mu| b \sqrt{\sigma^2 - \xi^2}}{\omega_T}
(-\sin\omega_T t, \cos\omega_T t, 0) \, . 
\label{momentum}
\ee 
This result is only in part equivalent to the classical one of Eq. 
(\ref{zeq})   because here the instantaneous position value 
$\lps {\bf x} \rps$ is 
always zero.

\section{Quantum dynamics}
\label{Quantum}

\subsection{Effective Hamiltonian}
The quantum dynamics is more properly addressed by transforming the original 
equations into a spin reference frame rotating at the bias field frequency. 
Thus the wave function $| \Psi^{\rm R} \rps$ in the rotating frame is written as
$|\Psi^{\rm R} \rps =R_z(-\omega_T t)|\Psi\rps= exp(\frac{i}{\hbar}s_z  
\omega_T t)|\Psi \rps \, ,$
where $R_z(\vartheta)$ is the rotation around the $z$ axis by an angle 
$\vartheta$, and $| \Psi \rps$ is
the laboratory frame wave function.
Then the Schr\"odinger equation for $| \Psi^{\rm R}\rps$ becomes
\be
\begin{split}
i \hbar \partial_t |\Psi^{\rm R} \rps = & i \hbar \partial_t R_z(-\omega_T t) 
| \Psi \rps =  \\
&[ H^\prime(t) -\hbar \omega_T s_z  ] 
|\Psi^{\rm R} \rps \,=H^{\rm R}| \Psi^{\rm R} \rps ,
\end{split}
\label{schr2}
\ee
where $H^\prime(t)=R_z(-\omega_T t) H(t) 
R_z(\omega_T t)$ is time-dependent and 
\be
H^{\rm R} = H^\prime(t) - \hbar\omega_T s_{z}=  \frac{{\bf p}^2}{2m} + m g z - 
\frac{\mu}{s} {\bf s}\cdot {\bf B^{\rm R}}({\bf x},t) \, , 
\label{Hamrot}
\ee
with
$ 
{\bf B^{\rm R}}({\bf x},t)= [B_0 + b (x\cos\omega_T t + y\sin\omega_T t)] 
\hat{x} + b (y\cos\omega_T t - x \sin\omega_T t) \hat{y} 
+ (-2 b z + 
\omega_T s/\mu
) \hat{z} \, .
$
${\bf B^{\rm R}}$ is the magnetic field in the spin rotating 
frame, a constant bias field and a rotating inhomogeneous field.
This effective magnetic field  
${\bf B^{\rm R}}({\bf x},t) = B^{\rm R}_x \hat{x} + B^{\rm R}_y \hat{y} 
+ B^{\rm R}_z \hat{z}$ identifies the position dependent angles 
$\vartheta$ and $\varphi$ as
\be
\vartheta =
\arctan\frac{\sqrt{(B^{\rm R}_x)^{2} 
+ (B^{\rm R}_y)^2}}{B^{\rm R}_z} \, , \quad 
\varphi   =
\arctan{\frac{B^{\rm R}_{y}}{B^{\rm R}_{x}}} \, .
\label{angledef}
\ee
and $ B^{\rm R} =  \sqrt{(B^{\rm R}_x)^{2} 
+ (B^{\rm R}_y)^{2}+(B^{\rm  R}_z)^{2}}$.
\subsection{Local basis}
It is useful to
introduce a coordinate-dependent spin basis $\{|\chi_m ({\bf x},t)\rps
\}$ such that \be \frac{{\bf s} \cdot {\bf B^{\rm R}}({\bf
x},t)}{B^{\rm R}} |\chi_m ({\bf x},t)\rps = m |\chi_m({\bf x},t) \rps
\, \quad \text{for $-j \leq m \leq j$ } \, .
\label{localbasis}
\ee
The local basis vectors in which the $z-$axis coincides with
the magnetic field in the same point can be given in terms of
the angles $\vartheta$ and $\varphi$ through the rotation operator
$M(\varphi,\vartheta)$ as follows
\be
 | \chi_m (\vartheta,\varphi)\rps = M(\varphi,\vartheta)|j,k \rps=  
e^{-\frac{i}{\hbar} \varphi s_z}
e^{-\frac{i}{\hbar} \vartheta s_y}
| j,k \rps \, .
\label{basisalgebric}
\ee
With the total wave function expanded as
$
|\Psi^{\rm R}\rps = \sum_{m=-j}^j \psi^m({\bf x},t) \, |\chi_m ({\bf
x},t)\rps \, ,
$
the Hamiltonian of Eq.  (\ref{Hamrot}) becomes
\be
\begin{split}
H^{\rm R} = &\frac{{\bf p}^2}{2m} + m g z - 
\frac{\mu}{s}  B^{\rm R}({\bf x}) s_z + \\ 
 &\frac{1}{2m} \left\{
2 {\bf A} \cdot {\bf p} + {\bf p}({\bf A}) + {\bf A}^2
\right\} + {\cal V}
\end{split}
\label{Hamtot}
\ee
with 
\be
\begin{split}
{\bf A}  &= -(s_z \cos\vartheta - s_x \sin\vartheta) 
\nabla \varphi  - s_y \nabla \vartheta 
\, , \\
{\cal V} &= -(s_z \cos\vartheta - s_x \sin\vartheta)\partial_t \varphi -
s_y\partial_t \vartheta \, .
\end{split}
\label{nonAdTs}
\ee
Appendix \ref{spin} contains details useful to derive the functions in 
Eq. (\ref{nonAdTs}).
In Eq.  (\ref{Hamtot}) ${\bf A}$ and ${\cal V}$ represent pseudopotentials
connected to the Lorentz-like and electric-like kinds of forces introduced 
in Ref. \cite{aharonov}.  The functions
$\varphi({\bf x},t)$ and $\vartheta({\bf x},t)$ depend on the
effective magnetic field geometry as stated in Eqs.  (\ref{angledef}). 
Berry's geometric terms appear in the quantum Hamiltonian through these
angles.  By direct computation the following
relations are derived: 
\be
\begin{split}
\nabla \vartheta  =
\frac{\left[ {\bf B^{\rm R}} \wedge ({\bf B^{\rm R}}
 \wedge \nabla {\bf B^{\rm R}}) \right]_z}
{(B^{\rm R})^2 \sqrt{(B^{\rm R})^2 - (B^{\rm R}_z)^{2}}} \, &, \ \
 \nabla \varphi = 
 \frac{\left[ {\bf B^{\rm R}}
 \wedge \nabla {\bf B^{\rm R}} \right]_z}
{(B^{\rm R})^2 - (B^{\rm R}_z)^{2}} \, , \\
\partial_t \vartheta  =
\frac{\left[ {\bf B^{\rm R}} \wedge ({\bf B^{\rm R}}
 \wedge \partial_t {\bf B^{\rm R}}) \right]_z}
{(B^{\rm R})^2 \sqrt{(B^{\rm R})^2 - (B^{\rm R}_z)^{2}}} \, &, \ \
 \partial_t \varphi = 
 \frac{\left[ {\bf B^{\rm R}}
 \wedge \partial_t {\bf B^{\rm R}} \right]_z}
{(B^{\rm R})^2 - (B^{\rm R}_z)^{2}} \, . 
\end{split}
\label{dephitheta}
\ee
These terms, invariant with the modulus of the magnetic
field, are geometric fields depending only on the force
lines of the magnetic field, i.e., the field geometry.
Their explicit form is given in Appendix \ref{thetaphiexpl}.

Notice that, starting from the Hamiltonian (\ref{genHam}), which is linear in the spin 
operators, the nonadiabatic terms give rise to a dynamics with  
a quadratic dependence in the spins. We may write $H^{\rm R}$ 
in a form that put this in more evidence
\be
H^{\rm R} = \frac{{\bf p}^2}{2m} + m g z +
\sum_i h^i s_i + \sum_{j k} g^{j k} s_j s_k \, ,
\label{betterHamtot}
\ee
where the indices $(i,j,k)$ run on $(x,y,z)$, and
the spin coefficients are
\be
\begin{split}
h^x &= 
\frac{1}{2m} [{\bf p} \sin\vartheta \nabla \varphi + 
\sin\vartheta \nabla \varphi {\bf p} ] + \sin\vartheta
\partial_t \varphi
\, , \\
h^y &=  -\frac{1}{2 m}
[{\bf p}  \nabla \vartheta +  \nabla \vartheta{\bf p}] -
 \partial_t \vartheta
\, , \\
h^z &= - \frac{\mu}{s}  B^{\rm R}
- \frac{1}{2m}
[{\bf p} \cos\vartheta  \nabla \varphi + 
\cos\vartheta  \nabla \varphi {\bf p}] -
\cos\vartheta \partial_t \varphi
\, , \\
g^{xx}& = \frac{1}{2m} \sin^2\vartheta \vert \nabla \varphi \vert^2
\, , \\
g^{yy}& = \frac{1}{2m}  \vert \nabla \vartheta \vert^2
\, , \\
g^{zz}& = \frac{1}{2m} \cos^2\vartheta \vert \nabla \varphi \vert^2
\, , \\
g^{xz}& = g^{zx} =- \frac{1}{2m} \sin\vartheta \cos\vartheta \vert 
\nabla \varphi \vert^2
\,  .
\end{split}
\label{geomfields}
\ee
and $g^{jk} = 0$ otherwise. Let us recall that all the operators
$h_i,g_{jk}$  are hermitian ones.

\section{ Beyond the adiabatic approximation}
\label{Beyond}
\subsection{ Effective spin dynamics}
 Since the exact solution of the Schr\"odinger Eq. (\ref{schr2}) is an
 impracticable task, the spin dynamics will be taken into account in
 an effective way by resorting to a  time-dependent variational
 principle (TDVP) \cite{TDVP}.  The TDVP procedure allows to reduce 
the system quantum dynamics to a semiclassical Hamiltonian form. 
  This  procedure was introduced for studying the low-lying
 collective states in nuclei \cite{mandel}, but was later shown 
to provide a valid approximation also for the one particle 
Schr\"odinger equation. Within this procedure, whose details are shown 
in Appendix \ref{TDVP}, we choose a suitable trial state of the form
\be \psi({\bf x},t) =
\exp[-i {\bf {\cal{S}}}(t)/\hbar] \Psi({\bf x},t) |j,\tau (t)\rps \,
\label{trialstate}
\ee
which will be subjected to the weaker form of the Schr\"odinger
equation embodied into TDVP, i.e. Eq. (\ref{prima}) in Appendix \ref{TDVP}. 
Here $ \Psi({\bf x},t)$ and $|j,\tau (t)\rps$ take in account for the
center-of-mass motion and spin dynamics, respectively.  ${\bf {\cal{S}}}(t)$ 
is  an effective action for the spin variables.
By carrying out the variational procedure on the trial wave function 
we  derive the classical
equations of motion for the expectation-values of the spin-operators
$s_j$ on the spin component of the dynamical trial state $|j,\tau
(t)\rps$. A key point in this variational procedure is the 
parametrization of
the spin variables in terms of  coherent atomic states. 
These latter have the physical significance of angular momentum 
states produced by a classical source \cite{mandel}. 
They depend on a complex parameter $\tau$ and are defined as
\be
|j,\tau \rps = \frac{1}{[1+|\tau|^2]^{j}} 
\sum^j_{m=-j} \left[ \binom{2j}{(j+m)} 
\right]^{1/2} \tau^{j+m} |j,m\rps \, ,
\label{cs1}
\ee
where $|j,n\rangle$ are the spin basis with the quantization axis 
taken along the direction of the local field. 
These states, analogous to the coherent states of the 
electromagnetic field, are defined within 
a subspace determined by the angular momentum $j$.  Within this 
sub-space each state, completely defined by the complex number $\tau$, 
is mapped onto the direction of a vector on a sphere by a projective 
transformation \cite{TDVP}. In our case this  vector  identifies  the 
orientation of a 
classical spin with respect to a local frame having the $z$ axis along 
the local magnetic field  ${\bf B}^{\rm R}({\bf x},t)$. 
This property can be understood    by computing the expectation values 
of the spin components on $|j,\tau \rps$. By keeping in mind the  
parametrization $\tau =- e^{-i \varphi^\prime} \tan (\vartheta^\prime/2)$ we find 
\be
\begin{split}
{\cal{S}}_{\rm x} = \lps j,\tau | s_x | j,\tau \rps & =  j \hbar 
\sin\vartheta^\prime\cos\varphi^\prime \, ,\\
{\cal{S}}_{\rm y} = \lps j,\tau | s_y | j,\tau \rps & =  j \hbar 
\sin\vartheta^\prime\sin\varphi^\prime \, ,\\
{\cal{S}}_{\rm z} = \lps j,\tau | s_z | j,\tau \rps & =  j  \hbar 
\cos \vartheta^\prime \, .
\end{split}
\label{spmz1}
\ee
where $\vartheta^\prime$ and $\varphi^\prime$ are the angles between 
the classical spin and the local magnetic field. 
The detail of the spin dynamics derivation are contained in the Appendix \ref{TDVP}. 
Their ruling equations are generated by the classical Hamiltonian
of Eq. (\ref{clnlsh}).  \par 
Let us focus on the center-of-mass motion described by the wave function  
$\Psi({\bf x},t)$ as in Eq. (\ref{wf}). 
The trapping potential obtained 
by the application of TDVP procedure as from Eq.~(\ref{Action}) can be expanded 
in a power series of the displacement coordinates around the trap center 
by keeping  only the harmonic terms. As a matter of fact these terms depend
on the quantity $\sigma(t)$, with $\sigma(t)$ given by
 \be
 \sigma(t) =  \frac{{\cal{S}}_{\rm z}}{s} \, .
\label{sigma}
 \ee
that coincide with the definition introduced previously within the
adiabatic approximation.  Since the evolution of this quantity is much
slower than the bias-frequency $\omega_T$ ( $\omega_z \ << \omega_T$),
its time dependence is maintained in the ruling equations even after
averaging over the short time scale of the bias field time dependent
terms as done in order to arrive to Eq.  (\ref{Sered2}).  As a
consequence the wave function solution of (\ref{Sered2}) can be
written as \be \Phi ({\bf x},t,\sigma) = \sum_{\{{\bf n}\}} c_{{\bf
n}} {\cal E}_{{\bf n}}(t) \Phi_{{\bf n}} ({\bf x},\sigma) \, ,
\label{solwg}
\ee
where the vector index means ${\bf n}=(n_1,n_2,n_3)$ along the three 
orthogonal directions, and the constants 
$c_{{\bf n}}$ are determined by the atomic initial conditions. 
The functions $\Phi_{{\bf n}} ({\bf x},\sigma)$ are the eigenfunctions of the
three-dimensional harmonic oscillator with eigenvalues
\be
\begin{split}
E_{{\bf n}} (\sigma) =& U_0 + \frac{1}{2 m} \left(
 \frac{\sigma \mu b}{ \beta \omega_T}
 \right)^2
+ \\
&\hbar \omega_{r} (\sigma) (n_1+n_2 +1)
+ \hbar \omega_{z} (\sigma) (n_3 + \frac{1}{2}) \, ,
\end{split}
\label{eigv}
\ee
and ${\cal E}_{{\bf n}}(t) = \exp [i \gamma_{\bf n} (t)/\hbar - i \int^t_0 
dt E_{{\bf n}} (\sigma(t))/\hbar] $ embodies also a  geometric phase
\be
\gamma_{\bf n} (t) = i \hbar \int^{\sigma(t)}_{\sigma(0)} d \sigma 
\left[ 
\int d {\bf x} \bar{\Phi}_{{\bf n}} ({\bf x},\sigma) 
\frac{\partial}{\partial \sigma} \Phi_{{\bf n}} ({\bf x},\sigma)
\right].
\label{gemph}
\ee 
The  parameter  $\sigma (t)$ entering into
the equations of motion for the atomic 
center of mass is actually a dynamical degree of freedom whose evolution is 
generated by the classical spin dynamics. 
Thus, the center of mass motion and the spin dynamics interact the one with the 
other  and they 
must be  simultaneously integrated. Let us stress that the dynamics we have 
just found, is the classical canonical counterpart of that one generated by 
the full quantum Hamiltonian written in Eq. (\ref{betterHamtot}).

\subsection{Ground state configuration}
In order  to find the non-adiabatic corrections to
the ground state solution (\ref{wf}),  we assume that this 
solution is well represented also  if we keep the lowest order in the
expression of the Hamiltonian of Eq.~(\ref{clnlsh}), i.e., the first order
in $(x,y,\zeta)$ appearing in  $H^R$.  This means that the Hamiltonian 
parameters of  Eq.  (\ref{clnlsh}), the
classical form of Eq.  (\ref{betterHamtot}), can be computed as an
average on the adiabatic-like ground state solution of Eq.  (\ref{wf})
of the approximated operators $h^i,g^{jk}$.  Therefore at the lowest 
order of approximation 
the terms in Eq. (\ref{geomfields}) result 
\[
\begin{split}
h^x &\simeq \frac{1}{m\rho \beta} [-\sin (\omega_T t) p_x + \cos
(\omega_T t) p_y] \, , \\
h^y &\simeq  
-\frac{1}{m \rho \beta^2}\{
\eta [\cos (\omega_T t) p_x + \sin (\omega_T t) p_y ] +
2 p_z \}
\, , \\
h^z &\simeq - \frac{\mu b \rho}{s} \beta - \frac{\eta}{m\rho\beta}
[-\sin (\omega_T t) p_x + \cos (\omega_T t) p_y ] \, , \\
g^{xx}& \simeq 
\frac{1}{2m \rho^2 \beta^2} 
\, , \\
g^{yy}& \simeq 
\frac{3+\beta^2}{2m \rho^2 \beta^4}  
\, , \\
g^{zz}& \simeq 
\frac{\eta^2}{2m \rho^2 \beta^2}  
\, , \\
g^{xz}& \simeq 
- \frac{\eta}{2m \rho^2 \beta^2}  
\, .
\end{split}
\]
Recalling the expectation value of the momentum ${\bf p}$ given by 
Eq. (\ref{momentum}),  up to the order $1/\rho$ we have
\[
\begin{split}
\lps h^x_{(0)} \rps  &\simeq \frac{2\beta  }
{(1+\eta^{2}) } \frac{\omega_{0,r}^{2}}{\omega_T }\, , \\
\lps h^z_{(0)} \rps  &\simeq 
- \beta \frac{\mu B_{0}}{s} 
- \frac{2\eta \beta } {(1+\eta^{2}) }\frac{ \omega_{0,r}^{2}}{ \omega_T}
 \, , \\
\end{split}
\]
and 0 otherwise. The corresponding classical spin Hamiltonian is, 
apart a constant,
\[
{\cal H }({\cal{S}}_x,{\cal{S}}_y ,{\cal{S}}_z) = \lps h^x_{(0)} \rps 
{\cal{S}}_x + \lps h^z_{(0)} \rps {\cal{S}}_z \, ,
\]
whose equations of motion results
\be
\begin{split}
\dot{\cal{S}}_x  &= - \lps h^z_{(0)} \rps {\cal{S}}_y \, ,\\
\dot{\cal{S}}_y  &= - \lps h^x_{(0)} \rps {\cal{S}}_z + \lps h^z_{(0)} \rps 
{\cal{S}}_x \, ,\\
\dot{\cal{S}}_z  &=\ \lps h^x_{(0)} \rps {\cal{S}}_y  \, . 
\end{split}
\ee
By setting $\dot{\cal{S}}_i = 0$ with $(i=x,y,z)$  we determine the ground 
state configuration
\be
\begin{split}
({\cal{S}}^0_x,{\cal{S}}^0_y,{\cal{S}}^0_z) = (&\pm\frac{\lps
h^x_{(0)} \rps}{\lps h^z_{(0)} \rps} \frac{\hbar j} {\sqrt{1+\lps
h^x_{(0)} \rps^2/ \lps h^z_{(0)} \rps^2}}, 0 , \, \\
&\pm \frac{\hbar j}
{\sqrt{1+\lps h^x_{(0)} \rps^2/ \lps h^z_{(0)} \rps^2}} ).  
\end{split}
\label{spinew}
\ee 
The ${\cal{\bf S}}^{0}$ solution, with the spin aligned to
the local magnetic field, provides a correction to the adiabatic
approximation discussed before.  

\subsection{Effective classical dynamics} 
The center of mass motion is described by the wave-function 
$\Psi ({\bf x},t)= \Phi ({\bf x},t,\sigma) {\cal E}({\bf x},t,\sigma)$ 
introduced in (\ref{trialstate}). 
The exponential factor
${\cal E}({\bf x},t,\sigma)
$
has been  
defined in  Eq. (\ref{expfact})  and  
$\Phi ({\bf x},t,\sigma)$ satisfies the
time-dependent Schr\"odinger equation (\ref{Sered2}) for the $3$-dimensional 
harmonic oscillator which Hamiltonian is $H={\bf p}^2 / (2m) +
U_h({\bf x},\sigma(t))$. In terms of the frequencies 
(\ref{corrfreqs}) and of the equilibrium atomic position (\ref{height}),
the time-dependent harmonic potential has the form
$U_h({\bf x},\sigma(t))= m [\omega^2_r (\sigma(t)) (x^2+y^2) + \omega^2_z 
(\sigma(t))(z-h(\sigma(t)))^2]/2$.

Upon introducing the center of mass position ${\bf R} = \lps \Psi| {\bf x} |
\Psi \rps$ and momentum ${\bf P} = \lps \Psi| {\bf p} |\Psi \rps$, the
following classical equations of motion are easily derived
\be
\begin{split}
\frac{d{\bf R}}{dt}&= \frac{{\bf P}}{m} \, , \\
\frac{d{\bf P}}{dt}&= - \nabla_{\bf R} U_h({\bf R},\sigma(t)) 
- \frac{d{\bf \Delta P}}{dt}
\label{comdyn} 
\end{split}
\ee
where ${\bf \Delta P} = \int d {\bf x} |\Phi({\bf x},t)|^2 
\nabla w({\bf x},t)$, and $\sigma(t)$ is defined by 
Eq. (\ref{sigma}).  Now  we introduce a further factorization of the kind  
 $\lps \Psi|{\cal O}({\bf x},t) {\bf p}|\Psi \rps 
\approx {\cal O}({\bf R},t) {\bf P}$, 
where ${\cal O}({\bf x},t){\bf p}$ stands
for the first three among the operators appearing into Eqs. (\ref{geomfields})
expanded in a power series of ${\bf x}$ and ${\bf p}$ up to the second
order~\cite{notef}. Then by 
considering only the linear terms in spin variables appearing into the 
classical spin Hamiltonian (\ref{clnlsh}) we can write 
\be
\begin{split}
{\cal H }({\cal{S}}_{\rm x}, {\cal{S}}_{\rm y}, {\cal{S}}_{\rm z}) &=
\sum_i \lps h^i \rps {\cal{S}}_i \, ,
\end{split}
\label{clnlshA}
\ee
where the time-dependent coefficients are implicit 
in  the center of mass wave function $\Psi({\bf x},t)$ as
given in (\ref{cHc}). 
Thus in Appendix \ref{TDVP} we derive the following equation of motion 
for the classical spin 

\be
\frac{d{\bf{\cal{S}}}}{dt} = {\bf{\cal{B}}}(t) \wedge  {\bf {\cal{S}}} \, .
\label{spindyn}
\ee
where the  magnetic field  ${\bf{\cal{B}}}$ is the sum of the real  one 
plus some  fictitious terms having originated from  
the geometric forces with components 
\be
\begin{split}
{\cal{B}}_x (t) &=\frac{\left\{ {\bf B^{\rm R}}
 \wedge \left[ \left( \frac{{\bf P}}{m} \cdot \nabla
+ \partial_t  \right){\bf B^{\rm R}} \right]
 \right\}_z}
{B^{\rm R}\sqrt{(B^{\rm R}_x)^2 + (B^{\rm R}_y)^{2}}} \, , \\
{\cal{B}}_y (t) &=  - \frac{\left\{ {\bf B^{\rm R}} 
\wedge \left[{\bf B^{\rm R}}
 \wedge \left[\left(
\frac{{\bf P}}{m} \cdot  \nabla  + \partial_t\right)
 {\bf B^{\rm R}}\right] \right] \right\}_z}
{(B^{\rm R})^2 \sqrt{(B^{\rm R}_x)^2 + (B^{\rm R}_y)^{2}}} \, , \\
{\cal{B}}_z (t) &= - \frac{\mu }{s} B^{\rm R} 
- \frac{B^{\rm R}_z\left\{ {\bf B^{\rm R}}
 \wedge \left[ \left( \frac{{\bf P}}{m} \cdot \nabla
+ \partial_t  \right){\bf B^{\rm R}} \right]
 \right\}_z}
{B^{\rm R}[(B^{\rm R}_x)^2 + (B^{\rm R}_y)^{2}]}\, .
\end{split}
\label{geometric}
\ee
Thus, the two equations systems (\ref{comdyn}) and (\ref{spindyn}) 
form a closed system to be simultaneously integrated.

\section{Numerical simulations}
\label{Numerical}
\begin{figure}[ht]
\includegraphics[scale=0.7]{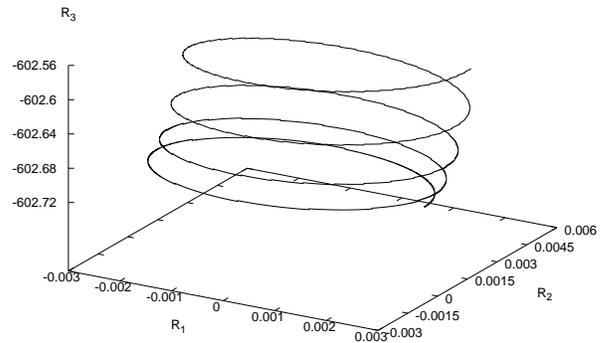}
\caption{Plot of a $3-$D trajectory  originated by initial conditions 
close to those corresponding to the 
micromotion.  The numbers on the axes express the
coordinates in $\mu$m. This trajectory corresponds to $5$ times the 
$2 \pi/\omega_T$ period. It is not closed, while  a stable micromotion orbit 
corresponds to a closed motion. The number on the axes express the coordinates in $\mu$m. }
\label{fig6}
\end{figure}

We have numerically integrated the set of equations (\ref{comdyn}) and 
(\ref{spindyn}) by means of a Runge-Kutta algorithm. The set of 
parameters chosen, i.e.   $B_0= 4.\ 10^{-4}$T, $b=0.18$T/m, and 
$\omega_T=2 \pi \ 10^{4}$s$^{-1}$,  correspond to those used in 
TOP experiments exploring the rubidium micromotion~\cite{muller00,cr00}.
The simulations allowed to recover the atomic micromotion, representing
periodic closed orbits. The micromotion was investigated 
through the classical equations of motion (\ref{comeom}) and also 
through the improved system of equations of  Eqs. (\ref{comdyn}) and 
(\ref{spindyn}). Similar results were obtained for the center of 
mass motion. In both approaches we observed a strong 
dependence on the initial conditions, that  for the classical 
center of mass variables are given by  Eqs. (\ref{height}) and 
(\ref{momentum}).
For the spin variable the classical condition corresponds to the spin 
aligned along the local ${\bf B}^{inst}$ magnetic field, while the 
quantum mechanical solution requires the atom to be in  an eigenstate of 
the spin operator along the local magnetic field. A modification of
the initial conditions from those required for the 
atomic micromotion, for instance a shift of $100\ \mu$m 
along the $z-$axis, produced the open trajectories shown in 
Fig.~\ref{fig6}. We noticed also a strong dependence on the initial 
condition for the atomic spin.
We also verified that within the parameters used here which approximately 
match  those corresponding to the experimental set up 
of Ref.~\cite{muller00}, the correction to the adiabatic 
approximation expressed by Eq. (\ref{spinew}) are not quite relevant.
We verified numerically that the spin projection  along the 
effective magnetic field ${\bf B}^{inst}$  given by Eq. (\ref{Binst}),
is well conserved while the spin projection along the real magnetic 
field ${\bf B}$  evidences time dependent oscillations, 
as already stated by Ref. \cite{schmiedmayer}.

\begin{figure}[ht]
\includegraphics[scale=0.4]{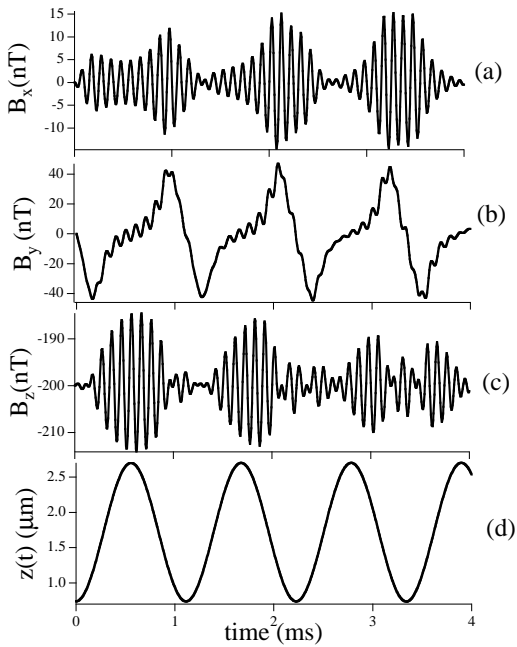}
\caption{In (a), (b) and (c) plots of the  ${\cal{B}}_x$ , 
${\cal{B}}_y$, and ${\cal{B}}_z$ 
geometric fields, in units of $1.\times 10^{-9}$T,
as a function of time in ms for atomic motion within a TOP trap with a 
RF bias field of 200 nT. In(d) the $z-$position, in $\mu$m, of the 
atomic center of mass 
traced by integrating the effective equations (\ref{comdyn}). Instead the 
integration of the classical equations of motion 
(\ref{comeom}) displays an unstable trajectory.}
\label{fig8}
\end{figure}

We have  explored a different region of parameter
values, where we expect  the adiabatic approximation to break down. 
While the adiabaticity is certainly not fulfilled   if $\omega_z,\omega_r 
\sim \omega_L$,
a further source of failure  for  this approximation  rests in the 
intensity of the geometric magnetic fields of  Eqs.
(\ref{geometric})  being of the same order 
of magnitude of  the applied real ones. This occurs, for example,  at bias 
field intensities  of the order of $B_0= 2.\ 10^{-7}$T.
Notice that for this weak rotating RF field, the trap oscillation frequencies of Eq. 
(\ref{frequencies}) are large enough to sustain the atoms against 
gravity. However the radius $\rho$ of the circle of death of Eq. 
(\ref{par1}) becomes comparable to the radius $r$ of the micromotion 
orbit.
Fig. \ref{fig8}  (a),(b) and (c) show the components of ${\bf {\cal{B}}}$ as a function 
of time obtained with initial 
conditions very close to those of  a micromotion orbit.
For this set of parameters
it is interesting to make a comparison between  the atom dynamics generated 
by the classical equation of motion and  the effective improved equations of 
motion. The latter give rise to a stable motion, shown in 
Fig. \ref{fig8}(d) traced by integrating the effective equations 
(\ref{comdyn}) and 
(\ref{spindyn}).  Under the same initial conditions the classical equations of motion 
generate an unstable trajectory  with the atoms conserving initially a 
constant height $z=
0.73 \mu$m, and then after several milliseconds escaping from the trap.
The more stable character of the effective equations solution in respect to
the classical ones is made evident by 
comparing the spin projection along ${\bf B}^{inst}$
in both cases. By numerically integrating the effective equations
(\ref{comdyn}) and (\ref{spindyn}), we found oscillations of 
${\bf {\cal S}} \cdot {\bf B}^{inst}/
|{\bf {\cal S}}| |{\bf B}^{inst}| $ near the stable value $1$.  
On the contrary, by integrating the classical equations (\ref{comeom}), we
found spin flip that causes the condensate escape from the trap.

\begin{figure}[ht]
\includegraphics[scale=0.7]{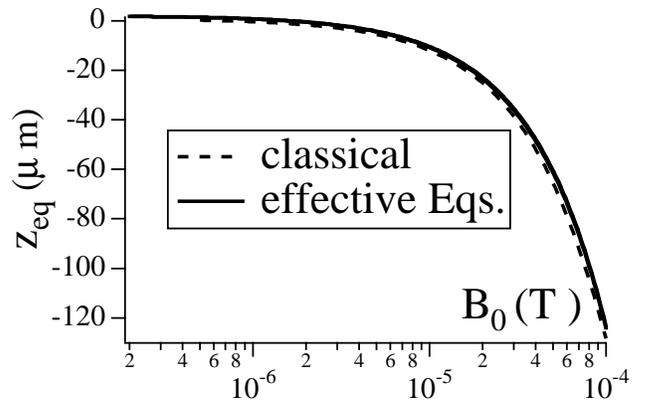}
\caption{Comparison between the equilibrium positions $z_{\rm eq}$
of rubidium atoms within the TOP trap  versus the $B_{\rm o}$ 
RF field as 
predicted by the classical solution and by the effective equations 
(\ref{comdyn}) and (\ref{spindyn}) at a given value of the quadrupole 
field gradient $b=0.18$T/m. For $B_{0} < 5.\times10^{-7}$T, the classical
model does not leads to a stable orbit.}
\label{fig4}
\end{figure}

The important role played by the terms originated by the 
non adiabatic approximation appears very clearly when we compared the 
atomic equilibrium within the TOP $z_{\rm eq}$ as derived by the classical solution 
to that predicted by the effective Eqs. (\ref{comdyn}) and 
(\ref{spindyn}). That comparison is shown  in Fig. \ref{fig4} for a 
fixed quadrupole field $b=0.18$T/m and a RF rotating field $B_{\rm 
0}$ between $1.\times 10^{-4}$T and  $2.\times 10^{-7}$T.  At large values 
of $B_{\rm 0}$ the values $z_{\rm eq}$ predicted by the classical 
solution and the improved one coincide. Instead the two values are 
different at small values of $B_{\rm 0}$ because the two solutions predict 
different equilibrium positions.  Finally, for $B_{0} < 5.\times10^{-7}$T,
classically the atoms are not suspended against gravity, while the
effective equations predict a stable equilibrium position.

\section{ Conclusions}
In this paper we have analyzed the motion of neutral atoms 
within TOP magnetic traps. We have considered the approximate 
classical equations of motion describing such system and 
revisited  the fast degrees of freedom motion at the forcing frequency 
$\omega_T$ known as atomic micromotion.
In such a motion non-adiabatic effects and geometric fields
are absent.
Within the scenario of adiabatic 
approximation, we have analyzed the atomic quantum motion by taking into 
account  the lowest frequencies terms embodied into the time-dependent 
potential.
Within this approximation, the center-of-mass motion resulting harmonic,
we have calculated the harmonic trapping frequences and  
recovered the quantum counterpart of the classical micromotion.
Addressing the problem within a pure quantum 
context, as a consequence of the presence
of inhomogeneous magnetic fields geometric magnetic fields appear quite 
naturally.
These geometric fields are responsible for a misalignment of the atomic
spin with   respect to the local magnetic field and then for non-adiabatic 
effects.
Within this framework, we have derived an effective classical dynamics in which
these  geometric fields are explicitly embodied. The atomic motion
results by the coupling of a quantum harmonic motion, governing the 
center of mass, and an effective nonlinear spin dynamics driven by both the 
local 
magnetic field and the geometric ones.\\
\indent The numerical simulations performed for the parameters of standard 
experimental set-ups  have shown that the adiabatic approximation is well
suited. On the other hand, by reducing the intensity of the bias field, 
non-adiabatic effects show up, because the geometric
field becomes  more relevant and cause misalignment of the spins around ${\bf 
B}^{inst}$.\\
\indent Another relevant facet concerns the  sensitivity to the initial 
condition of the trap equations. 
The projection of the atomic magnetic
moment on the local field ${\bf B}^{inst}$ is a conserved quantity,  
an adiabatic integral. The initial condition of ${\bf S}\cdot{\bf B}^{inst}$ 
chosen for a given simulation identifies a given dynamical evolution, 
its value  remaining conserved during atomic motion.
On the other hand, the equilibrium height is determined by the value of this
quantity [see Eq. (\ref{height})], so that the projection of the atomic 
magnetic moment on the local field determines the cloud  equilibrium height.
Now, it is a fact that, for a given geometry of magnetic fields within a trap, 
such a  height results  to be  independent on  the initial 
preparation condition. Thus, since the  dynamical equations do not 
select  by itself  any  special value of this  quantity,  
one could argue, in order to explain some experimental features of the 
BEC clouds \cite{muller00,cr00},   that a possible modification of the actual spin 
projection on 
${\bf B}^{inst}$ might  by involved in  some steps of the preparation of the
relative  Bose-Einstein condensate.
Further attention should also be devoted to the non-linear interaction within
the condensate.

\section{acknowledgments}
This work has progressed through the constant collaboration of 
several members of the BEC group in Pisa, M. Anderlini, D. Ciampini. 
R. Mannella, O. Morsch, J.H. 
M\"uller,  and through useful discussions with M. Mintchev.

\appendix
\section{About the su($2$) algebra}
\label{spin}
The su($2$) algebra is defined starting from the angular momentum 
generators $s_x,s_y,s_z$ and their commutation relations. Using the 
standard definitions for the raising and lowering operators, we derive 
the following relations for the derivatives of the operator 
$M(\phi,\theta)$ defined in Eq. (\ref{basisalgebric})
\be
\begin{split}
 M^\dagger(\varphi,\vartheta) \partial_\varphi M(\varphi,\vartheta) =& 
	-\frac{i}{\hbar} (s_z \cos\vartheta - s_x \sin\vartheta) \, , \\
 M^\dagger(\varphi,\vartheta) \partial_\vartheta M(\varphi,\vartheta) =& 
	- \frac{i}{\hbar} s_y \, , \\
 M^\dagger(\varphi,\vartheta) \partial^2_{\varphi \varphi} M(\varphi,\vartheta)=& 
	- \frac{1}{\hbar^2} (s_z \cos\vartheta - s_x \sin\vartheta)^2 \, , \\
 M^\dagger(\varphi,\vartheta) \partial^2_{\vartheta \vartheta} M(\varphi,
\vartheta) =& 
	- \frac{1}{\hbar^2} s^2_y \, , \\
 M^\dagger(\varphi,\vartheta) \partial^2_{\varphi \vartheta} M(\varphi,\vartheta)
 =& 
	- \frac{1}{\hbar^2} (s_z \cos\vartheta - s_x \sin\vartheta) s_y\, , \\
 M^\dagger(\varphi,\vartheta) s_x M(\varphi,\vartheta) =& 
	s_x \cos\vartheta\cos\varphi  
- s_y \sin\varphi + \\
& s_z \sin\vartheta \cos\varphi \, , \\
M^\dagger(\varphi,\vartheta) s_y M(\varphi,\vartheta) =& 
	s_x \cos\vartheta\sin\varphi  
+ s_y \cos\varphi + \\
&s_z \sin\vartheta\sin\varphi\, .
\end{split}
\label{average}
\ee

\section{Geometric terms}
\label{thetaphiexpl}
For a cylindric TOP trap with magnetic field as defined in 
(\ref{ciltop}), the geometric fields result
\be
\begin{split}
\nabla \vartheta &= \frac{b}{(B^{\rm R})^2 \sqrt{(B^{\rm R}_x)^2 +
(B^{\rm R}_y)^{2}}}
\begin{bmatrix} 
B^{\rm R}_z \, b (x + \rho\cos\omega_T t) \\ 
B^{\rm R}_z \, b (y + \rho\sin\omega_T t) \\ 
2[ (B^{\rm R}_x)^{2} + (B^{\rm R}_y)^2]
\end{bmatrix} \, , \\ 
 \nabla \varphi &= \frac{b^2}{(B^{\rm R}_x)^{2} + (B^{\rm R}_y)^2}
\begin{bmatrix}  
-y - \rho\sin\omega_T t \\ 
 x + \rho\cos\omega_T t\\ 
0   
\end{bmatrix} \, , \\ 
\partial_t \vartheta &= \frac{\omega_T b^2 B^{\rm R}_z }
{(B^{\rm R})^2 \sqrt{(B^{\rm R}_x)^2 + (B^{\rm R}_y)^{2}}} \, \rho \,
(y \cos\omega_T t - x \sin\omega_T t)\, , \\ 
 \partial_t \varphi &= \omega_T \left( \frac{b \rho B^{\rm
R}_x}{(B^{\rm R}_x)^{2} + (B^{\rm R}_y)^2} -1 \right) \, , \\
\Delta \vartheta &=\frac{B^{\rm R}_z} {\sqrt{(B^{\rm R}_x)^2 +
(B^{\rm R}_y)^{2}}} \left(1 + 6 \frac{(B^{\rm R}_x)^2 + (B^{\rm
R}_y)^{2}} {(B^{\rm R})^2} \right) \, , \\
 &\nabla \varphi \nabla \vartheta = 0 \, , \ \ \Delta \varphi = 0\,. 
\end{split}
\label{dephitheta2} \ee

\section{TDVP approach}
\label{TDVP}
In our contest  the TDVP method 
structures the dynamical quantum-state describing the atomic motion, 
in terms of a trial state written, as in Eq.  (\ref{trialstate}), 
as the product of a time-dependent
phase factor $e^{-i S(t)/\hbar}$ times a spatial- and time-dependent
scalar wave function $\Psi({\bf x},t)$ times a time-dependent spinor
$| j, \tau(t)\rps$.  The time-dependent trial state $\psi({\bf x},t)$
is to be found in a self-consisting way.  By imposing the weaker form
of the Schr\"odinger equation
\be
\int d{\bf x}\, \psi^\dagger({\bf x},t) \left[  i\hbar \partial_t  
 - H^{\rm R}\right]\psi({\bf x},t)
=0 \, ,
\label{prima}
\ee
we get the effective action $S(t)$
\[
\begin{split}
S(t) &= \int^t_0 d t \, \\ 
&\int d{\bf x}\,  \lps j, \tau(t)| \Psi^\dagger({\bf x},t) 
\left[  i\hbar \partial_t  
 - H^{\rm R}\right]\Psi({\bf x},t) | j, \tau(t)\rps \, .
\end{split}
\]
By splitting the Hamiltonian as $H^{\rm R} = H^\sigma_{ad} +
\Delta H^\sigma$, where $H^\sigma_{ad}$ has been introduced in
(\ref{Hadiab}), we get \be
\begin{split}
S(t) = \int^t_0 d t \, 
\int d{\bf x}\, &
\Psi^\dagger({\bf x},t) \left[  i\hbar \partial_t  
 - H^\sigma_{ad} \right]\Psi({\bf x},t)
+ \\
\int^t_0 d t \, 
\int d{\bf x}\, &\lps j, \tau(t)| 
\left[  i\hbar \vert \Psi({\bf x},t) \vert^2 \partial_t  
 - \right. \\ & \left.
 \Psi^\dagger({\bf x},t) \Delta H^\sigma 
\Psi({\bf x},t) \right]  | j, \tau(t)\rps \, .
\end{split}
\label{Action}
\ee
By expanding the Hamiltonian $H^\sigma_{ad}$ in a power series of the 
displacement coordinates $(x,y,\zeta=z-h)$, we get the harmonic Hamiltonian with the 
time-dependent potential of Eq. (\ref{u2t}). Thus, by introducing the structure
$\Psi ({\bf x},t) $ of Eq. (\ref{wf}) 
for the atomic wave function we find the Schr\"odinger Eq.~(\ref{Sered}).  
After taking the time average of the latter equation on a short time 
$2 \pi / \omega_T$, we get the harmonic problem (\ref{Sered2})
of which the general solution $\Phi ({\bf x},t)$ is known.
Thus, the first term in the r.h.s. in Eq. (\ref{Action}) vanishes, 
and we obtain
\[
S(t) = 
\int^t_0 d t \,  
\lps j, \tau(t)|
\left[  i\hbar \partial_t  
 - \lps \Delta H^\sigma_{(2)} \rps \right] | j, \tau(t)\rps \, ,
\]
where $\lps \Delta H^\sigma_{(2)} \rps = \int d{\bf x}\,[
\Psi^\dagger({\bf x},t) \Delta H^\sigma_{(2)}\Psi({\bf x},t)]$ and 
$\Delta H^\sigma_{(2)}$ is obtained
by expanding $\Delta H^\sigma$ in a power series in ${\bf x}$ and
${\bf p}$ up to the second order. Explicitly we have
\[
\lps \Delta H^\sigma_{(2)} \rps = 
\sum_i \lps h^i_{(2)} \rps s_i + \sum_{j k} \lps g^{j k} _{(2)} 
\rps s_j s_k - \lps \sigma \mu  B^{\rm R}_{(2)} \rps \, ,
\]
with the obvious meaning of index ${(2)}$.

The Hamiltonian $\lps\Delta  H^\sigma_{(2)} \rps$ is built of su($2$)  
algebra generators acting on the time-dependent spin vector $| j, \tau(t)\rps$.  
We choose for $| j, \tau(t)\rps$ the components of the su(2) atomic coherent 
state $|\tau(t) \rps$ of $j$ representation.
As a support for this choice let us remember that, if the Hamiltonian 
$\lps \Delta  H^\sigma_{(2)} 
\rps$ was a closed dynamical algebra, {\it i.e.} a linear combination of the 
su($2$) generators, the solutions of the Schr\"odinger equation given by these 
coherent states would be exact \cite{TDVP}.
The equation of motion for the label $\tau (t)$ (and its complex conjugate), 
that involves the dynamical evolution of the state 
$\psi({\bf x},t)$, is obtained by stationarizing the effective action 
$S(t)$ 
\be
\begin{split}
\delta S = \delta &\left(
\int^t_0 d t \, 
\Bigg\{
 i \hbar  \, 
\lps j, \tau (t) | \left[ \dot{\tau} \frac{d}{d \tau}    +
 \dot{\bar{\tau}} \frac{d}{d \bar{\tau}}\right]  | j, \tau(t)  \rps  
 - 
\right. \Bigg. \\ &~~~~~~~~~~~~~~~~\Bigg. \left.
\lps j, \tau (t) | \lps \Delta H^\sigma_{(2)} \rps | j, \tau(t)  \rps
\Bigg\}
\right) = 0\, .
\end{split}
\label{dS}
\ee
 After boring algebra we get the
equations of motion
\[
\frac{d \tau}{dt} = \frac{(1+\vert \tau \vert^2)^2}{2 i \hbar j} 
\, \partial_{\bar{\tau}} {\cal H }(\tau,\bar{\tau} ) \, , \\
\]
where ${\cal H }(\tau,\bar{\tau} ) = \lps \tau | \lps \Delta  H^\sigma_{(2)}
 \rps | \tau \rps$ and $\bar{\tau}$ is the complex conjugate of $\tau$. 
These tricky equations can be cast in more familiar 
form by writing them in terms of the classical spin components
${\cal{S}}_{\rm x},{\cal{S}}_{\rm y},{\cal{S}}_{\rm z}$ 
introduced in the text. 
The dynamics of these spin components, apart terms irrelevant for the
equations of motion, is generated by the following Hamiltonian: 
\be
\begin{split}
{\cal H }({\cal{S}}_{\rm x}, {\cal{S}}_{\rm y}, {\cal{S}}_{\rm z}) &=
\sum_i \lps h^i_{(2)} \rps {\cal{S}}_i + 
\\ 
(1-\frac{1}{2j})& \sum_{i}
\lps g^{i i} _{(2)} \rps {\cal{S}}^2_i + 2 (1-\frac{1}{2j}) \lps
g^{xz} _{(2)}\rps {\cal{S}}_{\rm x} {\cal{S}}_{\rm z} \, ,
\end{split}
\label{clnlsh}
\ee
where the time-dependent coefficients are derived by
the center of mass wave function $\Psi({\bf x},t)$ as
\be
\begin{split}
\lps h^i_{(2)} \rps &= \int d{\bf x}\,[
\Psi^\dagger({\bf x},t) h^i_{(2)} \Psi({\bf x},t)]
\quad  {\rm for} \ i = x,y,z \, , \\
\lps g^{ij}_{(2)} \rps &= \int d{\bf x}\,[
\Psi^\dagger({\bf x},t) g^{ij}_{(2)} \Psi({\bf x},t)]
\quad  {\rm for} \ i,j = x,y,z \, ,
\end{split}
\label{cHc}
\ee
the operators $h^i$ and $g^{ij}$ have been given in (\ref{geomfields}).
Therefore we obtain a canonical system for the classical spin components 
whose dynamics is generated by the effective Hamiltonian ${\cal H }$, endowed
with the Poisson brackets $\{{\cal{S}}_{\rm x},{\cal{S}}_{\rm y}\} =
{\cal{S}}_{\rm z}$ and cyclic permutations.



\begin{thebibliography}{99}
%
\bibitem{wilczek} For an introduction see a supplement in Sakurai {\it
Modern Quantum Mechanics} (Addison Wesley, Boston, 1994). For a
review see A.~ Shapere and F.~Wilczek, {\it Geometric phases in
Physics} (World Scientific, Singapore, 1989).  For a  list of
experiments on Berry's phase see J.~Anandan, J.~Christiam, and
K.~Wakelik, Am. J. Phys. {\bf 65}, 180 (1997).

\bibitem{majorana} An early study of the non-adiabatic processes for the
atomic motion within an inhomogeneous magnetic field was made by
E.~Majorana, Nuovo Cimento {\bf 9}  44 (1932).

\bibitem{aharonov} Y.~Aharonov and A. Stern, Phys. Rev. Lett. {\bf 69}, 3593 (1992).

\bibitem{littlejohn} An additional velocity-dependent potential is derived in 
R.G.~Littlejohn and S. Weigert, Phys.  Rev.  A {\bf 48}, 924 (1993).



\bibitem{sun} C.-P.~Sun, M.-L. Ge, and Q. Xiao, Commun. Theor. Phys. {\bf 13} 
63 (1990);  C.-P.~Sun, M.-L. Ge, Phys. Rev. D {\bf 41}, 1349 (1990). 


\bibitem{ho} T.-L. Ho and V.B.~Shenoy, Phys. Rev. Lett. {\bf 77}, 
2595 (1996).

\bibitem{sukumar} C.V.~Sukumar and D.M.~Brink, Phys. Rev. A {\bf 56}, 
2451 (1997).


\bibitem{hinds00} E.A.~Hinds and C. Eberlein, Phys. Rev. A {\bf 61}, 
033614 (2000).

\bibitem{potvliege} R.M.~Potvliege and V.~Zehnl\'e, Phys. Rev. A {\bf 
63}, 025601 (2001).


\bibitem{schmiedmayer} J. Schmiedmayer and A. Scrinzi, Quantum
Semiclass.  Opt. {\bf 8}, 693 (1996); J.R. Anglin and J. Schmiedmayer,
arXiv:physics/0211062.





\bibitem{muller00}
J.H. M\"uller, O. Morsch, D. Ciampini, M. Anderlini, R. Mannella,
and E. Arimondo, Phys. Rev. Lett. {\bf 85}, 4454 (2000).

\bibitem{cr00} J.H.~M\"uller, O.~Morsch, D.~Ciampini, M.~Anderlini, R.~Mannella
and E.~Arimondo, C. R. Acad. Sci. Paris, {\bf t.2 IV}, 649 (2001).

\bibitem{petrich} W. Petrich, M.H.~Anderson, J.R.~Ensher, and E.A.~Cornell,
Phys.  Rev.  Lett.  {\bf 74}, 3352 (1995).

\bibitem{cornell95} M.H.~Anderson, J.R.~Ensher, M.R.~Matthews,
C.E.~Wieman, and E.A.~Cornell, Science {\bf 269}, 198 (1995).

\bibitem{hagley} E.W.~Hagley , L.~Deng, M.~Kozuma, J.~Wen,
K.~Helmerson, S.L.~Rolston, and W.D.~Phillips, Science {\bf 283}, 1706 (1999).

\bibitem{han} D.J.~Han, R.H.~Wynar, Ph.~Courteille, and D.J.~Heinzen,
Phys.  Rev.  A {\bf 57}, R4114 (1998).

\bibitem{anderson}B.P.~Anderson and M.A.~Kasevich, Phys.  Rev. A {\bf 59}, R938 (1999).

\bibitem{martin} J. L. Martin, C.R.~McKenzie, N.R.~Thomas,
J.C.~Sharpe, D.M.~Warrington, P.J.~Manson, W.J.~Sandle and
A.C.~Wilson, J. Phys.  B: Atom.  Mol.  Opt.  Phys.  {\bf 32}, 3065
(1999).

\bibitem{arlt} J. Arlt, O.~Marag\'o, E.~Hodby, S.A.~Hopkins,
G.~Hechenblaikner, S.~Webster, and C.J.~Foot, J. Phys.  B: Atom.  Mol. 
Opt.  Phys.  {\bf 32}, 5861 (1999).

\bibitem{GovStr}
S. Gov and S. Shtrikman, J. Appl. Phys. {\bf 86}, 2250 (1999).

\bibitem{note} The different experimental configurations are characterized by the 
components of the quadrupole gradients along the orthogonal axes 
appearing in 
Eq. (\ref{gradient}) and by the plane containing the rotating magnetic 
field of Eq. (\ref{rotating}). The geometry of the Pisa experiment 
with a triaxial TOP trap 
corresponds to $-b_x=-b_y=b_z/2=b$ and the $B_{0}$ field rotating in 
the horizontal plane\cite{jphysbpaper}.

\bibitem{jphysbpaper}J.H.~M\"uller, D.~Ciampini, O.~Morsch, G.~Smirne, 
M.~Fazzi, P.~Verkerk, F.~Fuso, and E.~Arimondo,
J. Phys. B: Atom. Mol. Opt. Phys. {\bf 33}, 4095 (2000).

\bibitem{Minogin} V. G. Minogin, J. A. Richmond, and G. I. Opat,
Phys. Rev. A {\bf 58}, 3138 (1998).

\bibitem{SPIE} D.S.~Hall, J.R.~Ensher, D.S.~Jin, M.R.~Matthews,
C.E.~Wieman, and E.A. Cornell,  Proc.  SPIE Int.  Soc.  Opt.  Eng. 
{\bf 3270}, 98 (1998).

\bibitem{levitron} M. V. Berry, Proc. R. Soc. London A {\bf 452}, 1207-1220 
(1996).

\bibitem{cook} R. J. Cook, D. G. Shankland, and A. L.
Wells, Phys.  Rev.  A, {\bf 31}, 564 (1985).

\bibitem{TDVP}
W. M. Zhang, D. H. Feng and R. Gilmore, Rev. Mod. Phys. {\bf 62}, 867 (1990);
T. Fukui and Y. Tsue, Prog.  Theor.  Phys.  {\bf 87}, 627 (1992);
 A. Montorsi and V. Penna, Phys.  Rev.  B, {\bf 55}, 8226 (1997).
 \bibitem{mandel} L. Mandel and E. Wolf, {\it Optical Coherence and 
Quantum Optics}, (Cambridge University, Cambridge, 1995).


\bibitem{notef} 
Since the center of mass motion is described by a 3D harmonic 
oscillator Hamiltonian, the center of mass wavefunction is given at 
all times by a 
Glauber coherent state $| \alpha \rps$  representing the  minimum 
indetermination state\cite{mandel}.  It is easy matter to verify that 
$\lps \alpha | (x p + p x)/2 |\alpha \rps = 
\lps \alpha | x |\alpha \rps \lps \alpha| p |\alpha \rps$.
Therefore the factorization applies to those states. 

\end{thebibliography}
\end{document}